\documentclass[twocolumn,secnumarabic,amssymb, amsmath, nofootinbib,tightenlines,
nobibnotes, aps, prl]{revtex4}
\usepackage{amsmath}%
\usepackage{longtable}%
\usepackage{bm}%
\expandafter\ifx\csname package@font\endcsname\relax\else
 \expandafter\expandafter
 \expandafter\usepackage
 \expandafter\expandafter
 \expandafter{\csname package@font\endcsname}%
\fi

\usepackage{graphicx}
\usepackage{graphicx}

\begin{document}

\title{ Skin Friction in Simple  Wall - Bounded Shear Flows in the Large Reynolds Number Limit.}

\author{Victor Yakhot}
\email{vy@bu.edu}
\affiliation{Department of  Mechanical Engineering, \\
Boston University, Boston, MA 02215}
\date{today}%


\begin{abstract}
\noindent    A global approach to analysis  of  fully developed turbulent flows in pipes/channels and zero pressure gradient boundary layers is proposed. A new dynamic definition of the boundary layer thickness $\delta(x)$, where $x$ is the distance to the plate origin,   is proposed.  The  Coles -  Fernholtz empirical correlation for skin friction  $\lambda=\frac{2\tau_{w}}{\rho U_{0}^{2}}\propto 1/\ln^{2}\delta(x)$  and  $\delta(x)\propto x/\ln^{2}(\frac{x}{x_{0}})$  are derived from  the Navier-Stokes equations in the limit $Re_{x}\rightarrow \infty$.  Here $\tau_{w}$ and $U_{0}$ are the wall shear stress and free stream velocity, respectively. The theory is formulated as an expansion in powers of a small dimensionless parameter $\frac{d\delta(x)}{dx}\rightarrow 0$ in the limit $x\rightarrow \infty$. 

 \end{abstract}

\maketitle

 
\noindent   The law of  variation of  skin friction with Reynolds number 
in  turbulent wall flows  is one of the oldest riddles  of physics of turbulence.  In addition to the difficulties associated with a general problem of strong  isotropic turbulence,
the presence of solid walls  is responsible for  appearance  of two  different characteristic velocities. The so called friction velocity,  reflecting properties of the near-wall sublayer,  is defined as $u_{*}^{2}=\nu|\frac{\partial U(y)}{\partial y}|_{wall}$,  so that  for the dimensionless distance to the wall $y_{+}=\frac{yu_{*}}{\nu}=O(1)$,   the ratio $U_{+}=U/u_{*}$ is independent upon Reynolds number. In fully developed pipe/channel flows,  the parameter $u_{*}$  can be expressed in terms of  a  prescribed constant pressure gradient (or gravity)  and the friction factor  relates mean velocity to  a driving force.  In the vicinity of the  centerline  ($y/H\approx 1$),  the  velocity  $U(y)\approx U_{cL}$  must be found as a solution to dynamic equations of motion.  In the zero-pressure-gradient boundary layers with externally prescribed free stream  velocity $U_{0}$,  this parameter is related to the shear stress at the plate and, in addtion,  to the boundary layer thickness $\delta(x)$, which in this case, depends upon distance to the origin  $x$.  It is the  interplay  of these two characteristic velocities  which makes theoretical evaluation of velocity profiles $U(y)$    a very difficult problem.

\noindent Since  the skin friction $\lambda=\frac{2\tau_{wall}}{\rho U^{2}}\approx (\frac{u_{*}}{\overline{U}})^{2}$,  all previous  calculations heavily relied on  a detailed knowledge of   theoretically  (and experimentally) uncertain function  $U(y)$  needed for calculation  of mean velocity $\overline{U}$.
The analysis of pipe/channel flows  is typically based on an   assumed scaling relation for velocity  represented in the "inner" and "outer" regions of the flow as 
$ U(y)=u_{*}f(y_{+})$ and $U(y)=U_{cL}-u_{0}g(\frac{y}{H})$, respectively  [1].
The parameters $u_{*}$ and $u_{0}$ are corresponding characteristic velocities.   Then, different matching conditions  applied   in the "overlap"  region lead  to different shapes of velocity profile $U(y)$.   
 To obtain  the functional form of $U(y)$ from a systematic  {\it local}   theory,  one must derive an expression for the  distribution of  the Reynolds stress $\tau_{x,y}(y)$ which is equivalent to solution of a proverbial "turbulence problem".  Therefore, at the present time,  scaling of  skin friction  with  Reynolds number remains   an unsolved problem.  
 
 \noindent In a recent paper, assuming the logarithmic velocity profile across  a  zero-pressure-gradient boundary layer,   Nagib et al [2] developed an asymptotic expansion,  leading to the so called Coles-Fernholtz relation [3]:
 
 \begin{equation}
 \lambda\propto \frac{1}{\ln^{2} Re_{\delta(x)}}
 \end{equation}

\noindent  widely accepted as an accurate large Reynolds number asymptotics. While this work is based on a solid mathematical analysis, its starting point, logarithmic profile, is an assumption not following   the Navier-Stokes equations.  

\noindent   In the present paper  we present  a simple global approach, leading to  the expression (1)  for  the skin friction  and $\delta(x)\propto x/\ln^{2}\frac{x}{x_{0}}$  for  the thickness  of  zero-pressure-gradient boundary layers  not relying upon any information about  local features of wall  flows.   
The theory is based on the  following concept:
Isotropic and homogeneous turbulence can be characterized by a {\it single dimensionless} parameter,  called Reynolds number:  $Re=u_{rms}L/\nu\rightarrow \infty$ where $u_{rms}^{2}=\overline{u^{2}}$ and $L$ is an integral scale at which energy is pumped into the system due to external forcing or large-scale instability.  Various renormalization procedures based on perturbation expansions in powers of this ("bare" ) Reynolds number,  led to  effective or renormalized, viscosity $\nu_{T}\approx u_{rms}L$, widely used in engineering turbulence modeling.   The main outcome of the method is   a reformulated perturbation  series  in powers of  "dressed" or renormalized Reynolds number $Re_{T}=u_{rms}L/\nu_{T}=O(1)$, for which no resummation method has been developed.  It is the lack of a small parameter  approaching  zero in the limit $Re\rightarrow \infty$ which makes this problem so hard.

\noindent In this respect,  the situation with  wall flows is at least as difficult and evaluation of the energy spectrum and scaling exponents of structure functions is an unsolved problem. 
 However, in this case, as $Re\rightarrow \infty$,  the {\it global dimensionless parameters}  $u_{*}/\overline{U}\rightarrow 0$ and $\frac{d\delta(x)}{dx}\rightarrow 0$  are small and can be used for construction of the well-behaved perturbation expansion leading to prediction of  global properties of wall flows.  This is the main goal of this paper.\\
 
\noindent  {\it Channel/Pipe flows.} First, we consider a steady fully developed flow between two infinite plates separated by a gap $y_{g}=2H$, so that $H\leq y \leq -H$.  (The centerline is at $y=0$).The flow is driven by the  pressure gradient $\frac{\partial p}{\partial x}=\frac{p(x+L)-p(x)}{L}=const$.  Using the Reynolds decomposition of velocity field ${\bf v}=U{\bf i}+{\bf u}$ where $\overline{{\bf v}}=U(y){\bf i}$, the Navier-Stokes equations for incompressible fluid ($\rho=const$, $\nabla\cdot {\bf u}=\nabla\cdot {\bf v}$) can be written as:

\begin{eqnarray}
\partial_{i}(u_{i}{\bf u})+U\partial_{x}{\bf u}+u_{y}\partial_{y}U{\bf i}=-\frac{\nabla p}{\rho}+{\bf i}\nu \partial^{2}_{y} U+\nu\nabla^{2}{\bf u}
\end{eqnarray}

\noindent and, since all derivatives $\partial_{x}\overline {\Psi}=\partial_{z}\overline {\Psi}=0$, where $\overline{\Psi}$ is the mean value of an arbitrary flow property $\Psi$  and $z$ is a coordinate in the span-wise direction, we have:

\begin{equation}
\partial_{x}p/\rho-\partial_{y}\tau_{x,y}=\nu\partial^{2}_{y}U
\end{equation}

\noindent where  the Reynolds stress $- \overline{u_{x}u_{y}}=\tau_{x,y}$.   
 The relation  (3) expresses the Reynolds stress in terms of an unknown velocity distribution $U(y)$. To close the  problem,  one  has to   write  a differential  equation  for $\tau_{x,y}$,  which involves a new unknown function,  for which one has  to derive  another equation and so on  {\it ad infinitum}. The procedure,  leading  to an infinite chain of partial differential equations,  can  easily be formally written down but is too hard to solve. Here we propose a global approach not relying upon information about local properties of the flow.
 
 \noindent Integrating (3) in the interval $0\leq y\leq H$ yields $H\partial_{x}p/\rho =-u_{*}^{2}=\nu\partial U|_{H}$ and denoting  $\tau_{xy}\equiv \tau$ and the centerline velocity $U_{cL}\equiv U(y=0)$, we derive:

$$
U(y)-U_{cL}=-\frac{u_{*}^{2}H}{2\nu}\frac{y^{2}}{H^{2}}-\frac{1}{\nu}\int_{0}^{y}\tau(y)dy
$$

\noindent Finally,  introducing dimensionless parameters  $Z=y/H=y_{+}/R_{*}$,  $y_{+}=yu_{*}/\nu$, $U_{+}=U/u_{*}$, $R_{*}=u_{*}H/\nu$ and $\tau_{x,y}= u_{*}^{2}\tau_{+}$, 
the equation valid for both  laminar and turbulent flows is readily derived:

\begin{eqnarray}
\overline{U}-U_{cL}=-\frac{u_{*}^{2}H}{6\nu}-\frac{1}{\nu H}\int_{0}^{H}dy\int_{0}^{y}\tau(y')dy'\nonumber \\=-\frac{u_{*}}{6}R_{*}-u_{*}R_{*}\int_{0}^{1}dZ\int_{0}^{Z}\tau_{+}(Z')dZ'
\end{eqnarray} 

\noindent  
To evaluate the integral (4) we need an expression for $\tau_{+}(y)$, which, at this time, is impossible to derive without generating an infinite chain of partial differential eqautions.  Instead, let us define a thicknes of sublayer $y=y_{sL}$, which combined with the expression (4), gives an exact magnitude of a global property $\overline{U}-U_{cL}$. In other words,  the integral in the right side of (4) is:

\begin{eqnarray}
\overline{U}-U_{cL}=-\frac{u_{*}}{6}R_{*}+
u_{*}R_{*}\int_{0}^{1-\frac{y_{sL}^{+}}{R_{*}}}dZ\int_{0}^{Z} Z'dZ' -I\nonumber \\
\approx -\frac{u_{*}y_{sL}^{+}}{2}-I
\end{eqnarray}

\noindent with 
$$
I=u_{*}R_{*}\int^{1}_{1-\frac{y_{sL}^{+}}{R_{*}}}dZ\int_{0}^{Z}\tau_{+}(Z')dZ' =u_{*}y_{sL}^{+}<\tau_{+}>
$$

\noindent  and  $<\tau_{+}><0$ denoting  the mean value of  dimensionless Reynolds stress in the sublayer $R_{*}-y_{sL}^{+}\leq y_{+}\leq R_{*}$.


\noindent Thus,

\begin{equation}
\overline{U}-U_{cL}=-\alpha u_{*}+O(1/R_{*})
\end{equation}

\noindent where $\alpha=y^{+}_{sL}(\frac{1}{2}+<\tau_{+}>)$. The formula (6) defines the  Taylor expansion in powers of a small parameter $u_{*}/\overline{U}$.  We can see that as $R_{*}\rightarrow \infty$,  the  dimensionless parameter $\psi=1-\frac{\overline{U}}{U_{cL}}\propto \frac{u_{*}}{U_{cL}}\rightarrow 0$ which reflects the fact that, with increase of the Reynolds number,  the velocity profile $U(y)$ flattens. This small parameter is crucial for the theory developed below. Based on  numerical and experimental data $y_{sL}^{+}\approx 30$, and $|<\tau_{+}>|\approx 0.4-0.45$  gives $\alpha \approx 4.$ (See, for example  Ref. [4]),. 
 
\noindent The formally exact relation (6)  has recently  been verified  by Zagarola et al [5] in experiments on the Princeton SuperPipe giving $\alpha\approx 4.3$ for $10^{5}\leq Re_{D} \leq 10^{7}$. Similar  result  can  be  obtained by integrating  the relation    $U_{cL}-U(y)=u_{*}F(\frac{y}{R})$  [6] in the interval $0\leq y\leq R$. This  gives $U_{cL}-\overline{U}=\alpha u_{*}$ where  $\alpha=\int_{0}^{1} F(x)dx$.
The  shape of the  function $F(x)$,  consistent with   logarithmic velocity distribution,  was used by Prandtl who, based on experimental information,  obtained  $\alpha \approx 3.75$. The later, probably more accurate,  measurements  gave $\alpha\approx 4.0$  (see  Ref. [7]  and references therein).

\noindent It follows from (6),   that  the skin friction in the pipe flow is equal to:
$
\lambda=8(\frac{u_{*}}{\overline{U}})^{2}=\frac{8}{\alpha^{2}}(\frac{U_{cL}}{\overline{U}}-1)^{2}
$

\noindent The predictions from this  relation   with $\frac{8}{\alpha^{2}}\approx 0.42$, are very close to   experimental data collected  from a  smooth pipe by McKeon et al  and from   honed and commercial rough pipes studied  by Schockling et al  and Langelansvik et al [8], respectively, which is close  $\alpha\approx 4.$ estimated above.

\noindent As  $Re\rightarrow \infty$,  the sublayer dominated by intermittent bursts of velocity derivatives,  dissipation and production,  can be considered   as a  low - Re turbulent flow with the mean velocity $U(y_{sL})\propto  u_{*}$ and the y-component of the fluctuating velocity $w(y_{sL})\propto  u_{*}$.  The  kinetic energy generated in the subalyer ($0\leq y_{+}\leq 15-30$)  is rapidly mixed and dissipated in the bulk.   It is clear that  in the large Reynolds number limit  $y_{sL} \rightarrow 0$,  the mean  energy flux through the  separating surface $(x, y_{sL}, z)$ is $\rho K(y_{sL})w(y_{sL})LW$   and   the energy balance can be written as :

\begin{equation}
\rho K(y_{sL})w(y_{sL})LW=O( u_{*}^{3}LW)=\rho \overline{{\cal E}}LWH
\end{equation}

\noindent  where 
$
\overline{\cal E}=\frac{1}{H} \int_{0}^{H}{\cal E}(y)dy=O(u_{*}^{3}/H)
$.

\noindent {\it Flat Plate Boundary Layer.}  We consider a flat plate $0\leq x\leq \infty$ and $y=0$. The freestream velocity of incoming flow is ${\bf U}_{0}=U_{0}{\bf i}$ and we are to analyze the Navier-Stokes -Prandtl equations in the boundary layer  approximation:
 
 \begin{equation}
 \frac{\partial U}{\partial x}+\frac{\partial V}{\partial y}=0
 \end{equation}

 \begin{equation}
 U\frac{\partial U}{\partial x}+V\frac{\partial U}{\partial y}=\frac{\partial}{\partial y}(\nu\frac{\partial U}{\partial y}+\tau_{ij})
 \end{equation}

\noindent As  $x\rightarrow \infty$ we,  assuming  self-similarity of the velocity profile write:
$U=U(\frac{y}{\delta(x)})\equiv U(\eta)$, $V=V(\frac{y}{\delta(x)})\equiv V(\eta)$ and $\tau_{x,y}=\tau_{x,y}(\frac{y}{\delta(x)})\equiv \tau_{x,y}(\eta)$ where the defined below  width of the boundary layer $\delta(x)$ {\it must be found from equations of motion}. The incompressibility constraint (8) gives:

\begin{equation}
V(x,y)=-\int_{0}^{y}\frac{\partial U(x,y')}{\partial x}dy'=\frac{\partial \delta(x)}{\partial x}\int_{0}^{\eta}\eta'\frac{d U(\eta')}{d\eta'}d\eta'
  \end{equation}




\noindent Integrating  (9)  over the interval $0\leq y\leq \infty$,  and introducing the `displacement thickness'  $\theta$ we, using (10),  express  the skin friction in terms of the boundary layer thickness  $\delta$:

\begin{equation}
\frac{d\theta}{dx}=\frac{\partial \delta(x)}{\partial x}\int_{0}^{\infty}\frac{U(\eta)}{U_{0}}(1-\frac{U(\eta)}{U_{0}})d\eta=\frac{u_{*}^{2}}{U_{0}^{2}}\propto \lambda
\end{equation}


 
\noindent  where $u_{*}^{2}=\nu \frac{\partial U}{\partial y}|_{0}$. Based on as yet unknown function $\delta(x)$, we  define an  averaged- over- the -boundary -layer  property $\Psi$ 
$\overline{\Psi}\approx \frac{1}{\delta}\int_{0}^{\delta}\Psi(y)dy=\int_{0}^{1}\Psi(\eta)d\eta$.  
 Since at  the edge of a boundary layer $y=\delta(x)$, the  velocity is $U=U(\delta(x))$ and kinetic energy $K=K(\delta(x))$, the familiar integral balance equations must be somewhat modified.  For example, integrating the differential energy balance equation:

$$
U\frac{\partial K}{\partial x}+V\frac{\partial K}{\partial y}=-\tau_{xy}\frac{\partial U}{\partial y}-{\cal E}+\frac{\partial}{\partial y}(\nu \frac{\partial K}{\partial y}+Q)
$$

\noindent in the interval $0\leq y\leq \delta(x)$ and recalling that

$$\int_{0}^{\delta}V\frac{\partial K}{\partial y} dy=V(\delta)K(\delta)-\int_{0}^{\delta}K\frac{\partial V}{\partial y} dy$$

\noindent  we, using an incompressibility constraint,  derive:

\begin{eqnarray}
\int_{0}^{\delta}\frac{d}{dx}K(x,y)U(x,y)dy+V(x, \delta)K(\delta)=\nonumber \\-\int_{0}^{\delta}\tau_{xy}\frac{\partial U}{\partial y}dy-\delta \overline{{\cal E}}+Q(\delta)
\end{eqnarray}

\noindent where $Q(\delta(x))=\overline{w(\delta)u_{i}^{2}(\delta)}$ is small.  In the limit $\delta\rightarrow\infty$, $V(x,\delta)\rightarrow 0$ and the relation (12) tends to a familiar energy balance (see for example Hinze [xx]).  With 
$V(x,\delta(x))=-\frac{d}{dx}(\delta \overline{U})+\frac{d\delta}{dx}U(\delta)$, directly following from (10), 
 the relation (12) takes a very simple form:

\begin{equation}
\frac{d}{dx}(\delta\overline{KU})-K(\delta)\frac{d}{dx}(\delta \overline{U})=\int_{0}^{\delta}\tau_{xy}\frac{\partial U}{\partial y}dy-\delta \overline{{\cal E}}+Q(\delta)
\end{equation}

\noindent Now, based on the results of a previous section, we introduce a new definition of a boundary layer thickness:

\begin{equation}
U_{0}-\overline{U}\approx \alpha u_{*};\hspace{1cm} \overline{K}\approx b u_{*}^{2};\hspace{1cm} K(\delta)\approx c u_{*}^{2}
\end{equation}

\noindent {\it Since $U_{0}> U(\delta)>\overline{U}$, then,   according to  proposition  (14),   $1-\frac{U(\delta)}{U_{0}}=\frac{\overline{U}}{U_{0}}-\frac{U(\delta)}{U_{0}}+\alpha \frac{u_{*}}{U_{0}}>0$.  
Therefore,  $U(\delta)-\overline{U}=O(u_{*})$,  which crucially differs from  a widely accepted ad hoc engineering definition of the boundary layer thickness $U(\delta)= 0.99 U_{0}$ implying $\psi=1-\overline{U}/U_{0}=O(1)$.   Let us demonstrate that  the anzatz (14),  combined with the energy balance (13),  leads  the well-known empirical relation $\lambda\propto \frac{1}{\ln^{2} Re_{\delta}}$.  } As follows from (14) and (11),   
$
\overline{KU}\approx \beta u_{*}^{2}(U_{0}-au_{*})
$  and:

\begin{equation}
\frac{d\delta}{dx}u_{*}^{2}U_{0}=O(\frac{u_{*}^{4}}{U_{0}})\ll u_{*}^{3}; \hspace{1cm} \frac{d\delta}{dx}u_{*}^{3}=O(\frac{u_{*}^{5}}{U_{0}^{2}})\ll \frac{u_{*}^{4}}{U_{0}};
\end{equation}

\noindent  It will become clear below that as $x\rightarrow \infty$,
$
\delta u_{*}^{2}\frac{du_{*}}{dx}=O(u_{*}^{5}/U_{0}^{2})
$ . Substituting the anzatz (14) into the energy balance (13) and  accounting for the  estimates  (15)  we, equating the terms of the same powers in $u_{*}$ and neglecting the  $O(u_{*}^{5}/U_{0}^{2})$ contributions,  derive $
\beta=c
$, 

$$
\int_{0}^{\delta}\tau_{xy}\frac{\partial U}{\partial y}dy=\delta \overline{{\cal E}}+Q(\delta)=O(u_{*}^{3})
$$

\noindent and

\begin{equation}
U_{0}\delta \frac{d u_{*}^{2}}{dx}=\alpha_{1}u_{*}^{3}\frac{d\delta}{dx}
\end{equation}

\noindent where $\alpha_{1}=a-\alpha<0$ is  an unknown  parameter of this global approach which must be obtained from a full local theory. It will become clear below that the balance (13) is possible only if $\alpha_{1}<0$.

\noindent It is easy to see  that  the expression: 

\begin{equation}
\lambda=2(\frac{u_{*}}{U_{0}})^{2}=\frac{\kappa}{\ln^{2} \delta}; \hspace{1cm} \lambda \propto \frac{d\delta}{dx}
\end{equation}

\noindent with $\kappa=8/\alpha_{1}^{2}$ is a solution to (16).  Indeed, integrating  (16)  and dividing the outcome by $U_{0}^{3}$, we obtain:

$$
\frac{\lambda}{2}=\frac{|\alpha_{1}|\sqrt{\kappa}}{4\sqrt{2}}\int \frac{d\lambda}{dx}dx=\nonumber \\
\frac{|\alpha_{1}|\sqrt{\kappa}}{4\sqrt{2}}\lambda
$$

\noindent      {\it This result shows that the anzatz (14) with $\lambda\propto \frac{1}{ \ln^{2} \delta}$ is a solution to the Navier-Stokes -Prandtl equations of motion.}

\noindent   Setting for a time being all proportionality coefficients equal to unity, we introducing   $\delta_{0}=\frac{\nu}{U_{0}}$, $Re_{\delta}=\frac{U_{0}\delta}{\nu}$ and $Re_{x}=\frac{U_{0} x}{\nu}$ solve the  the differential equations  (17) with the result:
$
Re_{\delta}[(\ln Re_{\delta})^{2}-2\ln \frac{Re_{\delta}}{e}]= Re_{x}
$ and,  as $Re_{\delta}\rightarrow \infty$, 

\begin{equation}
\delta(x)\rightarrow \frac{x}{\ln^{2} \frac{x}{\delta_{0}}}
\end{equation}



\noindent {\it Summary and discussion.} 
1.\ In this paper, based on the Navier-Stokes equations,  for a channel/pipe flow we derived the scaling relation (6), valid in pipe/channel flows the large-Reynolds number limit.  Known for many years, this formula was previously obtained from analysis of experimental data or assumed scaling  of  velocity profile $U(y)$.  While the linear scaling  with $u_{*}$ is an exact consequence of the Navier-Stokes equations, the magnitude of parameter $\alpha\approx 4.$ was estimated in this work from the near - wall data on the Reynolds stress. 

\noindent 2. \ This result led to a new dynamic definition of the boundary layer thickness given by (14).

 \noindent 3. \  The expression (17) for friction factor (skin friction)  was found as a solution to the Navier-Stokes-Prandtl  equations.

 \noindent 4.\ These results are accurate up to the $O(u_{*}/U_{0})$-corrections.


\noindent I would like to thank S. Bailey and J. Schumacher  for their  comments  and  invaluable 
help in comparing the results of this paper with experimental  and numerical data.  All ideas  involved in this work were discussed  in long and interesting conversations with  J. Schumacher,  K.R. Sreenivasan,  A. Smits, A. Polyakov, P. Monkiewitz   and A. Yakhot.

\end{document}